\documentclass{aa}

\usepackage{txfonts}
\usepackage{graphicx}

\begin{document}

   \title{Multi-color photometry of the Galactic globular cluster M75 = NGC 6864
   \thanks{Based on observations with the 1.3 m Warsaw telescope at Las Campanas
Observatory}.
   A new sensitive metallicity indicator and the position of the horizontal branch in UV}

      \author{V. Kravtsov\inst{1,2}
          \and
              G. Alca\'ino\inst{3}
          \and
              G. Marconi\inst{4}
           \and
              F. Alvarado\inst{3}
              }

\offprints{V. Kravtsov}

   \institute{Instituto de Astronom\'ia, Universidad Cat\'olica del Norte,
              Avenida Angamos 0610, Antofagasta, Chile\\
              \email{vkravtsov@ucn.cl}
            \and
              Sternberg Astronomical Institute, University Avenue 13,
              119899 Moscow, Russia\\
            \and
              Isaac Newton Institute of Chile, Ministerio de Educaci\'on de Chile,
              Casilla 8-9, Correo 9, Santiago, Chile\\
              \email{inewton@terra.cl, falvarad@eso.org}
            \and
              ESO - European Southern Observatory, Alonso de Cordova 3107, Vitacura,
              Santiago, Chile\\
              \email{gmarconi@eso.org}
             }

   \date{Received xxxxx / Accepted xxxxx}

   \abstract
{} {We carry out and analyze new multi-color photometry of the
Galactic globular cluster (GC) M75 in $UBVI$ and focus on the
brighter sequences of the color-magnitude diagram (CMD), with particular
emphasis on their location in $U$-based CMD. Specifically, we study
the level both of the horizontal (HB) and red giant branches (RGB)
relative to the main-sequence turnoff (TO) in the $U$ magnitude.}
{Along with the presented photometry of M75, we use our collection
of photometric data on GCs belonging to the metal-poor range,
[Fe/H]$_{ZW}<-1.1$ dex, obtained from observations with different
equipment, but calibrated by standard stars situated in the observed
cluster fields.} {We confirm our earlier finding, and extend it to a 
larger magnitude range. We demonstrate
that $\Delta U_{TO}^{BHB}$ expressing the difference in $U$
magnitude between the TO point and the level of the blue HB, near
its red boundary, of the metal-poor GCs observed with the EMMI
camera of the NTT/ESO telescope is about 0.4 - 0.5 mag smaller as
compared to GCs observed with the 100$\arcsec$ telescope and 1.3 m
Warsaw telescope of the Las Campanas Observatory. At the same time,
$\Delta U_{TO}^{RGB}$, the difference in $U$ magnitude between the
TO and RGB inflection (brightest) points, does not show such an
apparent dependence on the characteristics of $U$ filters used, but
it depends on cluster metallicity. We have shown, for the
first time, the dependence of the parameter $\Delta U_{TO}^{RGB}$ on
[Fe/H] and have estimated its analytical expression, by assuming a
linear relation between the parameter and metallicity. Its slope,
$\Delta U_{TO}^{RGB}$/$\Delta$[Fe/H]$\sim$1.2 mag/dex, is
approximately a factor of two steeper than that of the dependence of
the RGB bump position in the $V$ magnitude on metallicity. The
asymptotic giant branch (AGB) clump and features of the RGB
luminosity function (LF) of M75 are also discussed.} {}


   \maketitle

\section{Introduction}

Until recently, the southern Galactic globular cluster NGC 6864
(M75) has been among the least studied ones. Photographic
photometry in the GC by Harris (\cite{harris}), reaching the cluster
HB level, was the only CMD study of the globular.
Recent publications include an extensive study of M75 by
Catelan et al. (\cite{catelanetal}), based on deep multi-color
photometry reaching the cluster main sequence, the HST CMD of the
cluster obtained by Piotto et al. (\cite{piottoetal}), as well as a
study of a large number of the cluster variable stars (Corwin et al.
\cite{corwinetal}).

Using the present photometry of M75, particularly in the $U$
bandpass, we extend our results obtained
in a series of earlier studies of Galactic GCs. Carrying out 
photometry in GCs in the $U$ bandpass has been irregular. As a result, 
there is a deficit of
systematic studies of $U$-based CMDs over a wide luminosity range of
cluster stars, as well as over a wide metallicity range of GCs. As a
rule, $U$ photometry focuses on the distribution of cluster
stars along the HB in the ($U-V$) or ($U-B$) colors. However, deep
photometry in GCs in this bandpass can be valuable for other
applications as well.

Alca\'ino et al. (\cite{alcainoetal97b}b)
revealed, for the first time, a surprising disagreement
between the positions of the HB both in the $(U-B)$-$(B-V)$
two-color diagram and in the $U$-$(B-V)$ color-magnitude diagrams of
NGC6541 and M79, GCs with similar blue HB (BHB) morphology and 
metallicity. In particular, the disagreement of the BHB position in
the $U$ magnitude in the $U$-$(B-V)$ CMD of the two GCs was
around $\Delta U \sim$ 0.4 to 0.5 mag, while the slope of their RGBs
was the same within the error. The most pronounced difference in
the $U$ magnitude or in the ($U-B$) color between the BHB stars of
NGC 6541 and M79 occurred in the color range $\Delta(B-V)\sim$ 0.2
mag blue-ward of the red boundary of the BHB. The same kind of
discrepancy was also found comparing the corresponding diagrams of
M79 and M80 (Alca\'ino et al. \cite{alcainoetal98b}b).

Later, Momany et al. (\cite{momanyetal}) reported the so-called 
BHB red incursion through the RGB in the ($U-B$) color in the UV 
color-magnitude diagrams. The incursion is another manifestation 
of the same effect. Momany et al. (\cite{momanyetal}) have
investigated in detail the origin of this effect and have
convincingly showed that it is mainly due to the difference in
transmission curves of $U$ filters available at different
observatories. Specifically, transmission curves of some $U$ filters
used encompass the Bulmer jump, in conformity with the standard
$UBV$ photometric system, and using such filters one can realize
photometric bandpasses close to the standard $U$ one. The maxima of
their response curves fall between 3700 - 3750{\AA} or very close to
this range. In contrast to these filters, transmission curves of
other $U$ filters (hereafter referred to as $U\arcmin$ filters) 
are shifted blue ward of the Balmer jump, and the
resulting photometric band (hereafter $U\arcmin$ bandpass)
nearly does not include this jump. The $U\arcmin$ bandpass
response curves' maxima are approximately 150 - 200{\AA} or even
more blue-shifted, as compared to the location of the corresponding
maxima of the $U$ bandpasses. A number of examples of the normalized
transmission curves of $UV$ filters used at different observatories
are collected and presented in Momany et al. (\cite{momanyetal}).

The available information about the effect under consideration 
is usually expressed in the relative change of
the mutual position (either in $U$-based colors or in $U$ magnitude,
obtained with $U$ and $U\arcmin$ filters) between the BHB and RGB.
Moreover, it was obtained for GCs in a limited range of metallicity 
and with predominant BHB morphology. However, it is not clear 
(1) whether there are changes of the mutual positions in the $U$ and 
$U\arcmin$ not only
between the BHB and RGB, but also between them and the red HB (RHB) and
turnoff point of the same GC; (2) whether there is any dependence of
such changes on cluster metallicity. Therefore, we extend
our study to a wider range of $U$ (and $U\arcmin$) magnitudes of
cluster stars and a wider range of GC metallicity, as well as to
GCs exhibiting both the BHB and RHB.

\begin{table}[!t]
\begin{center}
\caption{Log of observations.} \label{tablog}
\begin{tabular}{cccccc}
\hline \hline \noalign{\smallskip}
$Night$ & $Exp.$ & $Filter$ & $Airmass$ & $Seeing$ & $Julian Day$ \\
%%\hline
    &   sec &   & & '' &   2451090.0$+$\\
\hline \noalign{\smallskip}
 $9/10$  &   40 &  V  &  1.201  & 1.1 &   6.58468\\
 $9/10$  &   60 &  B  &  1.207  & 1.1 &   6.58628\\
 $9/10$  &   90 &  U  &  1.215  & 1.2 &   6.58818\\
 $9/10$  &   40 &  I  &  1.225  & 1.0 &   6.59000\\
 $9/10$  &   40 &  V  &  1.246  & 1.1 &   6.59417\\
 $9/10$  &   60 &  B  &  1.253  & 1.1 &   6.59579\\
 $9/10$  &  120 &  U  &  1.262  & 1.2 &   6.59786\\
 $9/10$  &   40 &  I  &  1.276  & 1.0 &   6.59985\\
 $9/10$  &   40 &  V  &  1.299  & 1.1 &   6.60394\\
 $9/10$  &   60 &  B  &  1.308  & 1.2 &   6.60558\\
 $9/10$  &  120 &  U  &  1.319  & 1.3 &   6.60767\\
 $9/10$  &   40 &  I  &  1.335  & 1.0 &   6.60966\\

 $10/11$  &   40 &  V  &  1.247  & 1.1 &   7.59179\\
 $10/11$  &   60 &  B  &  1.255  & 1.1 &   7.59340\\
 $10/11$  &  120 &  U  &  1.264  & 1.2 &   7.59547\\
 $10/11$  &   40 &  I  &  1.278  & 1.1 &   7.59748\\
 $10/11$  &   40 &  V  &  1.286  & 1.1 &   7.59897\\
 $10/11$  &   60 &  B  &  1.295  & 1.2 &   7.60058\\
 $10/11$  &  120 &  U  &  1.305  & 1.3 &   7.60267\\
 $10/11$  &   40 &  I  &  1.320  & 1.1 &   7.60466\\
 $10/11$  &   40 &  V  &  1.330  & 1.1 &   7.60615\\
 $10/11$  &   60 &  B  &  1.339  & 1.2 &   7.60775\\
 $10/11$  &  120 &  U  &  1.351  & 1.3 &   7.60984\\
 $10/11$  &   40 &  I  &  1.368  & 1.2 &   7.61183\\

 $11/12$  &  120 &  V  &  1.234  & 1.0 &   8.58694\\
 $11/12$  &  180 &  B  &  1.248  & 1.1 &   8.58997\\
 $11/12$  &  360 &  U  &  1.265  & 1.2 &   8.59436\\
 $11/12$  &  120 &  I  &  1.296  & 1.0 &   8.59846\\
 $11/12$  &  120 &  V  &  1.332  & 1.1 &   8.60427\\
 $11/12$  &  180 &  B  &  1.349  & 1.2 &   8.60725\\
 $11/12$  &  360 &  U  &  1.373  & 1.2 &   8.61166\\
 $11/12$  &  120 &  I  &  1.413  & 1.2 &   8.61573\\
 $11/12$  &  120 &  V  &  1.434  & 1.2 &   8.61839\\
 $11/12$  &  180 &  B  &  1.456  & 1.2 &   8.62140\\
 $11/12$  &  360 &  U  &  1.485  & 1.3 &   8.62580\\
 $11/12$  &  120 &  I  &  1.537  & 1.1 &   8.62988\\
\hline \hline
\end{tabular}
\end{center}
\end{table}

\begin{table}[!t]
\begin{center}
\caption{Data on photoelectric standards} \label{tabstd}
\begin{tabular}{ccccccc}
\hline \hline
    Star &       X   &      Y     &       V     &      U-V    &     B-V  &      V-I\\
\hline
     A  &    680.8   &     799.1  &     13.650  &      0.090  &      0.720  &      0.840\\
        &            &            &     -0.028  &     -0.059  &     -0.034  &     -0.004\\
     B  &    880.8   &    1282.6  &     13.770  &      0.170  &      0.750  &      0.840\\
        &            &            &      0.006  &      0.005  &     -0.019  &     -0.007\\
     C  &    1060.6  &    1475.9  &     14.230  &      0.160  &      0.780  &      0.870\\
        &            &            &      0.012  &      0.035  &     -0.013  &     -0.023\\
     D  &     803.7  &    1501.6  &     14.240  &      0.480  &      1.010  &      1.040\\
        &            &            &     -0.023  &     -0.040  &      0.065  &      0.070\\
     G  &    1042.8  &    1540.3  &     15.210  &      0.280  &      0.800  &      0.820\\
        &            &            &      0.032  &      0.062  &      0.001  &     -0.036\\
\hline \hline
\end{tabular}
\end{center}
\end{table}

\section{Observations and data reduction}
\label{dataredu}

The observations were acquired on three nights,
October $9/10$, $10/11$ and $11/12$ 1998, with the 1.3 m Warsaw
telescope, Las Campanas Observatory, using a set of UBVI filters and
a $2048 \times 2048$ CCD camera with a gain $=3.8$ and a readout
noise of $5.5e^-$ rms. The array scale was $0\farcs417 {\rm
pixel}^{-1}$, giving a field of view of $14\arcmin \times 14\arcmin$.
The center of the measured field of NGC 6864 was approximately
$50\arcsec$ to the east and $10\arcsec$ to the south of the cluster
center. Flat-field, bias and dark frames were taken twice per night,
at the beginning and the end of each one. We took a total of 9
frames in $U$ (exposure time from 90 sec to 360 sec), 9 frames in
$B$ (60 sec to 180 sec), 9 frames in $V$ (40 sec to 120 sec), and 9
frames in $I$ (40 sec to 120 sec).  The average seeing estimated
from the observations was about $1\farcs0-1\farcs3$. Table
\ref{tablog} lists the log of the frames obtained.

The reductions of CCD photometry were performed at the Isaac Newton
Institute, and at the European Southern Observatory, ESO, Santiago,
Chile.  The stellar photometry was carried out separately for all
frames using {\sc daophot/allstar} (Stetson \cite{ste87},
\cite{ste91}). The program stars were detected and measured by
applying the usual procedures. To obtain the PSF, 20 to 30 stars in each
frame, bright but far from  saturation, were selected 
among those having no neighbors or defects within the PSF radius. We
find that among standard PSFs provided by DAOPHOT, the PENNY2
function enables us to handle aberrations, specific to individual
frames, most effectively. To calibrate our photometry, we relied on
photoelectric standards previously set up by Alvarado et al.
(\cite{awal90}) in the cluster field. It is the standard
approach used at the Isaac Newton Institute within the framework of
photometric studies of GCs. Table~\ref{tabstd} lists the photoelectric 
standards used in M75, their photoelecric magnitudes and
colors with the corresponding deviations, in the sense:
photoelectric values minus CCD ones. The formulae used in this study
to bring instrumental magnitudes and colors to the standard $UBVI$
photometric system are as follows:

$$V=v-0.055(\pm0.055)(b-v)+0.020(\pm0.041),$$
$$V=v-0.025(\pm0.038)(v-i)+0.019(\pm0.046),$$
$$U-B=1.099(\pm0.032)(u-b)-0.008(\pm0.028),$$
$$B-V=0.952(\pm0.012)(b-v)-0.032(\pm0.013),$$
$$V-I=0.962(\pm0.025)(v-i)+0.048(\pm0.025).$$

In our preliminary list, we retained only those stars that had at
least two measurements per night in each photometric band. We 
included in our final list only those stars for which at least 
one color index was determined. A total of
2681 stars were measured both in $V$ and $I$. Of them, 1547 and 659
stars have $B$ and $U$ magnitudes determined,
respectively\footnote{The results of our photometry are available
upon request.}. For stars of the brighter sequences, i.e. with $V <
18.0$, the r.m.s. errors are, on average, $0.028$ in $U$, $0.021$ in
$B$, $0.017$ in $V$, and $0.013$ in $I$.

Numerous stars measured by us in M75 are in common with the study by
Catelan et al. (\cite{catelanetal}). We found 1434 such stars. To
compare both photometries, however, we used the most reliable data,
namely those based on cross-identifications of the brighter stars
($V < 19.0$) in the less crowded parts of the cluster field. From a
sample of 621 stars retained, we have obtained the following
estimates of the mean differences between the two photometries, in the
sense Catelan et al. (\cite{catelanetal}) minus the present study:
$\delta V$ $=-0.058 (\pm0.087)$; $\delta (U-B)$  $=+0.011
(\pm0.085)$; $\delta (B-V)$ $=-0.108  (\pm0.076)$; $\delta (V-I)$
$=-0.018 (\pm0.079)$. The (B-V) color difference between the
photometries is larger than the corresponding differences
in other colors, which are quite small. This difference becomes
apparent at $(B-V) < 1.3$ and $V > 16.0$. Calibration of stellar
photometry by standard stars located in observed star fields should,
in principle, be more accurate than a calibration relying on
external standard stars. Although we use standard stars located in
the observed field of M75, the situation is not certain. The
number of the available standard stars within our field is limited
(5 stars), and among them there is no star in the blue part of the
color range $\Delta(B-V)$ covered by stars in the cluster CMD. For
this reason, we cannot exclude systematic effects in our calibration,
in particular in the color $(B-V)$.

  \begin{figure}
   \centering
   \includegraphics[angle=-90,width=6cm]{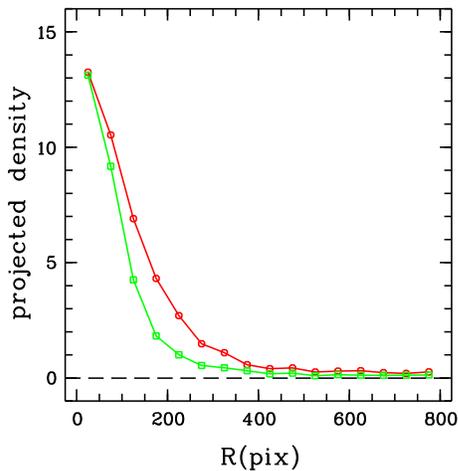}
      \caption{The dependence of the projected density of stars in the field of M75,
expressed (on a linear scale) as the number of stars per $10^3$ square pixels, as a
function of the radial distance from the center of the cluster; red
circles and green squares connected by lines of the corresponding
colors show the projected densities calculated for two limiting
magnitudes, $I_{lim}$=19.5 and in $V_{lim}$=19.0, respectively.}
         \label{projdens}
   \end{figure}

\section{The color-magnitude diagrams}

\subsection{General comments on the cluster and field star color-magnitude diagrams}

The field of M75 is observed to be contaminated by a number of 
field stars. To simultaneously obtain a larger sample of
cluster stars and to reduce the effect of the field stars on the
results of our photometry, we optimize the cluster-to-field star 
ratio in the analyzed sample of stars. We have determined the 
dependence of the projected density
of stars in the field of M75 as a function of the radial distance
from the center of the cluster. We show this dependence in
Fig.~\ref{projdens} for two limiting magnitudes, $I_{lim}$=19.5 (red
line and squares) and $V_{lim}$=19.0 (green line and squares). The
projected density is expressed as the number of stars per $10^3$
square pixels. The estimated densities converge near the cluster
center because of decreasing numbers of detected faint stars in the
central part due to crowding effects. However, at larger radial
distance, the sample of stars with fainter limiting magnitude
shows a higher projected density. The most
significant changes in the projected density of stars
occur within a radius $R <$ 500 pixels ($\sim$ 200\arcsec).
With increasing distance from the cluster center, the density
decreases much more gradually and asymptotically approaches its
constant value. This implies that at $R >$ 500 pixels, the 
projected density of cluster stars belonging to
the brighter sequences becomes comparable to that of the field
stars. Hence, we accept that $R\sim$ 500 pixels is the border radius
to sample a major fraction of the cluster population. It should be noted,
however, that this radius can vary for the cluster star belonging to
different sequences or magnitude ranges, since the cluster and field
stars show distinct distributions in the CMD. In particular, in the
region of the cluster BHB, there are no or a very small number of
field stars.

\begin{figure*}[t!!]
  \centering
  \resizebox{\hsize}{!}
 {\includegraphics[angle=-90]{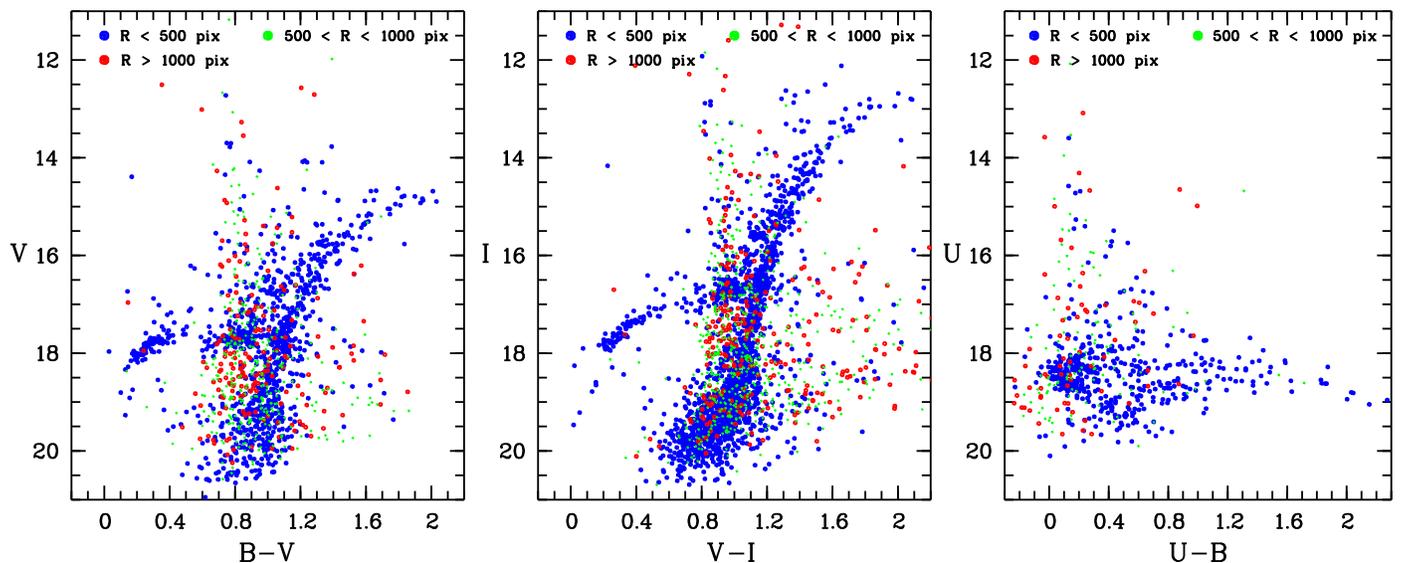}}
   \caption{The $V$-$(B-V)$, $I$-$(V-I)$ and $U$-$(U-B)$ CMDs for stars in
13$\arcmin$x13$\arcmin$ field of the Galactic GC M75. Blue symbols
denote stars with the projected distance $R <$ 500 pixels ($\sim$
200\arcsec) from the cluster center, while red ones show stars from
the outer part (with $R >$ 1000 pixels) of the observed field, the
bulk of the latter being the field stars; stars located at
intermediate distance ($500 < R <$ 1000 pixels) from the cluster
center are presented by small green dots.}
         \label{cmd}
   \end{figure*}

We give our photometric result for measured stars in a wide
field of M75 in three panels of Fig.~\ref{cmd} where the
$V$-$(B-V)$, $I$-$(V-I)$ and $U$-$(U-B)$ CMDs are shown. Blue 
dots correspond to stars confined within the region with radius
$R$ = 500 pixels, red circles show stars from the external
part (with $R >$ 1000 pixels) of the observer field. Both cover 
approximately equal areas. Stars located at
intermediate distances ($500 < R <$ 1000 pixels) from the cluster
center are shown with small green dots.

The right panel of of Fig.~\ref{cmd} shows the somewhat uncommon
appearance of the HB in the CMD (between $0.0 < U-B < 0.3$ and $18.0
< U < 19.0$) with the $U$-based color and $U$ magnitude, as compared
to CMDs with widely used colors and magnitudes.  The BHB and RHB
fall approximately in the same color range and appear as a clump,
the former being a brighter, nearly horizontal sequence of stars
within this clump, and the latter as a fainter and tilted one.

The number of cluster stars in the external part should already be
quite small, taking into account their radial distance from the
cluster center and the expected density of the cluster stars at this
distance. The real number of field stars in the CMDs
is not more than the number of the stars shown by red symbols in the
corresponding panels. One can see that the contamination by the
field stars of the BHB and the upper RGB is negligible. A somewhat
larger number of field stars can fall in the region of the AGB,
RHB, and RGB at the level of the HB and especially below it. The
blue boundary of the field stars in our CMDs in the registered
magnitude range is almost vertical and approximately corresponds to
the color of the blue boundary of the RHB. The CMD of the stars from
the outer part of our field is in good general agreement with the
CMD of the field stars around M75, obtained by Catelan et al.
(\cite{catelanetal}). We note that the limiting $V$ magnitude
achieved in our photometry is somewhat deeper than can be seen in
the $V$-$(B-V)$ diagram because of a shallower limit in the $B$
bandpass. In the $V$-$(V-I)$ diagram, that is not shown in
Fig.~\ref{cmd}, the limiting magnitude is very close to the
magnitude of the turnoff point occurring, according to Catelan et
al. (\cite{catelanetal}), at $V$ = 21.22.

While the demonstrated CMDs include all the
measured stars with $R <$ 500 pixels, in our subsequent analysis of
the photometry we have excluded stars located in the region with $R
<$ 40 pixels ($\sim$ 16\arcsec) where photometric accuracy decreases
appreciably.

\subsection{The asymptotic and red giant branches}

Our photometry of M75 yielded a fairly large sample of stars belonging
to the evolved sequences. This allowed us to address the main features
of the star distribution on the AGB and RGB.

The CMDs in Fig.~\ref{cmd} show that the number of stars
tracing the AGB, at least in its most populated lower part, in the
magnitude range $\Delta V \approx$ 1.5 mag makes it possible to
unambiguously  identify an important feature, the so-called AGB clump
at the base of the branch. It is seen at
$V\approx$16.8 and in the color range 0.95 $\leq (B-V) \leq$ 1.10 in
the $V$-$(B-V)$ diagram. In the $I$-$(V-I)$ diagram, the clump
is seen at $I\approx$15.6. It is more evident in the
latter diagram thanks to its narrower color range,
1.10$\leq(V-I)\leq$1.20.

The formation of the AGB clump, as well as the well-known RGB
bump, is caused by a slowing down of the rate of stellar evolution
along the given evolutionary sequence(s). For more details
concerning the nature of the clump and useful parameters deduced for
it from the CMD, as well as for more references related to the
subject, we refer, in particular, to Ferraro et al.
(\cite{ferraroetal}) and Sandquist \& Bolte (\cite{sanqbolte}). In the
present paper, we have been able to estimate one of the parameters,
namely the difference between the $V$-levels of the HB and the AGB
clump, $\Delta V_{HB}^{clump}$. Ferraro et al. (\cite{ferraroetal})
note that the estimations of the given parameter are available for
very limited number of GCs, since the AGB is poorly populated. Also,
to avoid any ambiguity in deducing this magnitude difference, they
argue to rely on the level of the zero-age HB (ZAHB) and point out
that "one might ideally define the ZAHB level by adopting the
magnitude of the lower envelope of the observed HB distribution in
the region with 0.2 $< B-V <$ 0.6", i.e. red-ward of the blue
boundary of the instability strip. We followed the same procedure
and found $V_{HB}$ = 17.82 $\pm$ 0.03. In turn, $V_{clump}$ = 16.75
$\pm$ 0.03 that leads to $\Delta V_{HB}^{clump}$ = 1.07 $\pm$ 0.06.

To obtain the RGB luminosity function (RGBLF) of M75, we
avoid contamination of the RGB by (1) stars
belonging to both the AGB and RHB, (2) stars showing large deviation
from the sequence's fiducial line due to photometric error or (3)
possible field stars that appear among the RGB stars on one CMD, but
are displaced from the RGB on another CMD. We used the
advantage of multi-color photometry and proceeded in the following
way. In each of the CMDs, $V$-$(B-V)$ and $I$-$(V-I)$, we fitted the
mean locus of the RGB with a polynomial using the corresponding
commands in the MIDAS system. We next linearized the RGB by
subtracting for each star the color of the mean locus at its
luminosity level from the star's color index. To construct the
RGBLF, we used only those stars that satisfied our selection
criterion: their deviations, $\delta(B-V)$ and $\delta(V-I)$, from
the mean locus in both colors simultaneously did not exceed $\pm$
0.06 mag. On the one hand, this value is close to mean error in the
colors of the fainter RGB stars, i.e. at the level of the HB and
below it, and somewhat larger than the mean error of the brighter RGB
stars. On the other hand, this conditional boundary
of the RGB separates the bulk of its stars from the majority of
stars belonging to the asymptotic and red horizontal branches.

\begin{figure}
  \centering
 \includegraphics[angle=-90,width=6cm]{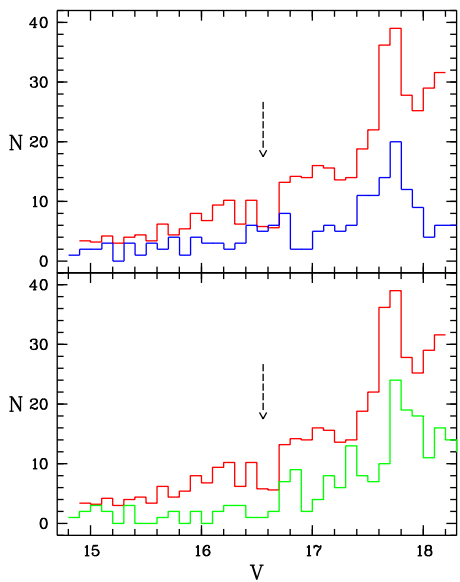}
   \caption{The RGBLF of M75 (blue line, upper panel) is compared with its counterpart
obtained for M80 (green line, lower panel) by Alcaino et al.
(\cite{alcainoetal98b}b), as well as with the generalized RGBLF (red
line) taken from Kravtsov (\cite{kravtsov}). The two latter LFs are
shown in the magnitude range of the former one, and they are shifted
along the $V$-axis until the coincidence of the bump position of all the
LFs. The generalized RGBLF is scaled down by a factor of 5 as
compared to the original one. Arrow indicates the gap present in the
RGBLFs of many GCs.}
         \label{rgblf}
   \end{figure}

The obtained RGBLF of M75 is shown by the blue line in the upper panel 
of Fig.~\ref{rgblf}. For a more reliable analysis, we
compare this LF with the analogous one (green line, lower panel)
obtained by Alca\'ino et al. (\cite{alcainoetal98b}b) in the GC M80.
Moreover, the generalized RGBLF (red line) from Kravtsov
(\cite{kravtsov}) is shown in both panels. It has been scaled
by a factor of 5 as compared to the original one. Also, this LF
and RGBLF of M80 have been reduced to the magnitude range of the
RGBLF of M75 by shifting them along the $V$-axis until reaching coincidence
of the bump position of all the LFs. The bump of the RGBLF of M75
is seen at $V_{bump}$ = 17.75 $\pm$ 0.05,
thanks to the sizable sample of RGB stars isolated in the region of
the bump. Taking into account the estimated level of the ZAHB at
$V_{HB}$ = 17.82 $\pm$ 0.03, we obtain a value for the magnitude
difference between the ZAHB and the bump $\Delta V_{HB}^{bump}$ =
0.07 $\pm$ 0.08. It is in good agreement with $\Delta V_{HB}^{bump}$ =
0.05 obtained by Catelan et al. (\cite{catelanetal}). This implies
virtually the same (within the uncertainty) cluster metallicity estimated
using the parameter $\Delta V_{HB}^{bump}$. Formally, our estimate
is a few hundredth dex more metal-poor than the one deduced by Catelan et al.
(\cite{catelanetal}), i.e., [Fe/H]$_{ZW}\approx -$1.33 dex in the scale
of Zinn \& West (\cite {zinnwest}).

The RGB bump is the only RGBLF feature that is widely accepted to be
physically real and in any GCs.
However, in the middle of 1980s one sometimes failed to detect
this feature, particularly in a number of the most
metal-poor GCs in which it was systematically less pronounced than
in more metal-rich GCs. As for other physically real features on
RGBLFs of GCs, their existence is still uncertain. Kravtsov
(\cite{kravtsov}) not only showed, for the first time, the
dependence of the RGB bump position in $V$ magnitude on GC
metallicity, but also studied the problem of the existence of other
possible RGB features. He used a generalized RGBLF
obtained from a sample of the LFs of the upper RGB parts of GCs
belonging to the metal-poor range. He concluded that the bump is
probably not the only real feature common to the RGB of
different GCs. The generalized RGBLF has revealed
three additional, statistically significant features that can be
related to the peculiarities of the evolution of stars along the RGB of
the clusters. At least one of them is especially pronounced. It
shows up as a quite significant gap on the generalized RGBLF, and
systematically appears in individual LFs as a depression or abrupt change in the LFs. Its %%@
statistical
significance in the generalized RGBLF was estimated and discussed in
Kravtsov (\cite{kravtsov}). Specifically, the
probability (P) that this feature in the generalized RGBLF is a
result of statistical fluctuation is P $<$ 0.001. The position in
$V$ magnitude of the gap under consideration is, on average, around
1.1 $\pm$ 0.1 above the bump. It is marked by rows in both panels of
Fig.~\ref{rgblf}. This feature was also noted by us in the RGBLF of
the Small Magellanic Cloud populous star cluster Kron 3 at $V \sim$
18.5 (Alca\'ino et al. \cite{alcainoetal96a}a) and in the RGBLF of
M80 at $V \sim$ 14.2 (Alca\'ino et al. \cite{alcainoetal98b}b). In
the magnitude range of the RGBLF of M75 it corresponds to $V \sim$
16.6. The RGBLF of M75 itself does not show any
significant gap at the given position. More exactly, the RGBLF
sampled with a smaller bin, $\delta V$ = 0.05, does show a
deep but narrow gap at this position. However, the general
behavior of the RGBLF of M75 in the magnitude range $\Delta V
\approx$ 2.0 mag above the bump is peculiar as compared to
that of the generalized LF, in the sense that they appear to be in
opposite "phase" to each other. Indeed, the former LF exhibits
two depressions where the latter one shows "normal" level and a small
local bump in the corresponding magnitude ranges (at 16.8
$< V <$ 17.4 and 16.0 $< V <$ 16.3, respectively), and vice versa,
i.e. at the location of the discussed gap we see
in the RGB of different GCs, the RGB of M75 rather has a local excess
of stars. Such a behavior of the RGBLF of M75 could explain
a deficit (if any) of the brighter RGB stars in M75, noted
earlier by Harris (\cite{harris}) and discussed and interpreted by
Catelan et al. (\cite{catelanetal}).

\subsection{The horizontal branch position in the CMDs with $U$ and $U\arcmin$ magnitude
axes}

As a first step in examining the position of the HB in the CMDs with
the UV magnitude axis, we compare the corresponding CMD of M75 with
the analogous deep diagrams of NGC 288 and NGC 6723, GCs whose
metallicities are confined in a limited range. The majoroty of
GCs, for which we have previously obtained deep photometry reaching
the TO in the UV range, were observed with the EMMI camera of the
NTT/ESO telescope in 1993. NGC 288 is among these clusters
(Alca\'ino et al. \cite{alilal97}c). In turn, photometry of NGC 6723
(Alca\'ino et al. \cite{alcainoetal99}), with limiting $U$ magnitude
below the TO as well, is based on observations gathered with the
100$\arcsec$ telescope of the Las Campanas Observatory. As was
established earlier and noted above, in Sect. 1, photometry
(above the TO in UV) with the same telescope and equipment in both
NGC 6541 (Alca\'ino et al. \cite{alcainoetal97b}b) and M80
(Alca\'ino et al. \cite{alcainoetal98b}b), on the one hand, and
photometry in the GC M79 (Kravtsov et al. \cite{kravtsovetal})
observed with the NTT, on the other hand, revealed disagreement in
the position of the BHB in the clusters' CMDs with UV magnitude
axis. For this reason, we use such diagrams for NGC 288 and NGC
6723, as templates of deep $U$-$(B-V)$ and $U\arcmin$-$(B-V)$ CMDs
of GCs with close metallicity and different HB morphology in order
to compare them to each other, as well as both to the corresponding
CMD of M75.

 \begin{figure}
  \centering
 \includegraphics[angle=-90,width=8cm]{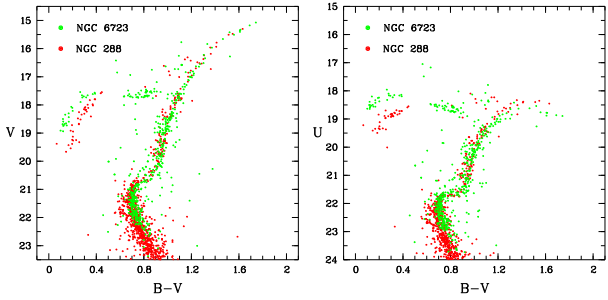}
   \caption{The superposition of deep $V$-$(B-V)$ and $U$($U\arcmin$)-$(B-V)$ diagrams of NGC
288 and NGC 6723, achieved by matching the CMDs in the region of the
turnoff point; both the magnitudes and colors of the original CMDs
are reduced to those of the corresponding CMDs of NGC 6864 (M75).}
         \label{superpos1}
   \end{figure}

 \begin{figure}
  \centering
 \includegraphics[angle=-90,width=8cm,clip]{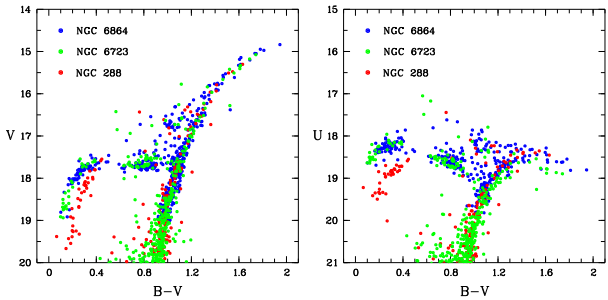}
   \caption{The superposition of upper parts of the $V$-$(B-V)$ and $U$($U\arcmin$)-$(B-V)$
diagrams of NGC 288, NGC 6723 and NGC 6864 (M75). }
         \label{superpos2}
   \end{figure}

\begin{table}
\begin{minipage}[t]{\columnwidth}
\caption{Color and magnitude offsets applied to the compared CMDs.}
\label{offset} \centering
\renewcommand{\footnoterule}{}
\begin{tabular}{lcccl}
\hline \hline Cluster  &$\Delta U$  & $\Delta V$ & $\Delta
(B-V)$\footnote{The CMDs are offset to the same reference points in
the system of the corresponding
CMDs of M75; the offsets are in the sense: \\
$(B-V)_{M75}-(B-V)_{GC}$ and $mag_{M75}-mag_{GC}$}& cluster photometry \\
\hline
 NGC 288  & 2.70  & 2.30& 0.220& Alca\'ino et al. (\cite{alilal97}c) \\
 NGC 1841  &-1.25  &    & 0.225& Alca\'ino et al. (\cite{alcainoetal96b}b)  \\
 NGC 1904  &2.00  &    & 0.300& Kravtsov et al. (\cite{kravtsovetal})  \\
 NGC 6397  &5.15  &    & 0.150&Alca\'ino et al. (\cite{alcainoetal97a}a)  \\
 NGC 6723  &2.50  & 2.15  & 0.130&  Alca\'ino et al. (\cite{alcainoetal99})  \\
 NGC 7099  &3.15  &     & 0.240&Alca\'ino et al. (\cite{alcainoetal98a}a)  \\
\hline
\end{tabular}

\end{minipage}

\end{table}

We superimposed the $V$-$(B-V)$ and
$U$($U\arcmin$)-$(B-V)$ CMDs of NGC 288 and NGC 6723. The
offsets along the luminosity axes were determined by best
coincidence of the diagrams around the turnoff points. The optimal
offsets along the color axes were achieved by the best matching the
positions of the turnoff points and the lower RGBs. We then
superimposed these matched CMDs and the same CMDs of M75. Since the
diagrams of this GC do not reach the TO, we relied on the optimal
coincidence, in the magnitudes and color, of the brighter sequences
of the CMDs of M75 and NGC 6723. Specifically, the mean levels of
the BHB and RHB were optimized in $V$ and $U$. The mean position in
the $(B-V)$ color of these branches and also of the RGB at the level
of the HB were used to define the offset in the color. As for the
offset along the $U$-axis between the CMDs of M75 and NGC 6723, the
validity of this procedure is justified by the realization, at the
100$\arcsec$ and 1.3 m Warsaw telescopes, of $U$ bandpass close to
the standard one, i.e. encompassing the Balmer jump. In particular,
according to data kindly provided by A. Udalski on a response curve
of the $U$ bandpass realized at the 1.3 m Warsaw telescope, the
wavelength of its maximum coincides with that of the standard curve.
Blue-ward of their maxima, the former curve shows a somewhat lower
level in comparison with the latter curve, and both curves show
nearly the same transmission in the region of the Bulmer jump.

The resulting superpositions of (1) the CMDs of NGC 288 and NGC 6723
in a wide luminosity range, from the main sequence to the RGB tip, and
of (2) the two GCs and M75 in the region of the brighter sequences,
are presented separately in Fig.~\ref{superpos1} and
Fig.~\ref{superpos2}. In both figures, the magnitudes
and colors of the original CMDs of NGC 288 and NGC 6723 are reduced
to those of the corresponding CMDs of M75. Their TO points are at
$V_{TO}$ = 21.20 $\pm$ 0.05, in good agreement with $V_{TO}$
= 21.22 $\pm$ 0.09 defined by Catelan et al. (\cite{catelanetal})
for M75. As seen from the $V$-$(B-V)$ plots, the RGBs of the three
GCs are virtually indistinguishable at the level of scatter and the
uncertainty of the superposition of the diagrams, and the slope of
the upper parts of the cluster RGBs is  identical. Also,
$V$-levels of the cluster BHBs near the blue edge of the instability
strip are the same within the uncertainty. It cannot be excluded that a
better coincidence of the BHBs and RHBs of NGC 6723 and M75 in the
$V$-$(B-V)$ diagram would be for slightly lower, by $\Delta V$ =
0.05 mag, positions of the CMD of NGC 6723 (in this case its upper
RGB would be slightly lower as well). However, the $U$-$(B-V)$ plots
show obvious differences between $U\arcmin$- and $U$-levels of
the BHB near the edge, i.e. between the BHB level of NGC 288, on the
one hand, and that of M75 and NGC 6723, on the other hand. In
agreement with our previous findings, this difference is of the
order of $\Delta U \sim$ 0.4 to 0.5 mag. At the same time, while the
RGBs of NGC 288 and M75 almost entirely coincide with each other and
show only small differences at their highest $U$-level (at
their inflection point), the RGB of NGC 6723 begins to
systematically deviate from the RGBs of these two clusters 
at $U <$ 19.5. The disagreement in $U$ magnitude between the RGBs
relative to the $U$ magnitudes of the corresponding TO points
becomes notable around the inflection points. In comparison with the
variation of the $U$-level of the BHB, this difference is mainly due
to cluster metallicity.

\begin{figure*}[t!!]
  \centering
  \resizebox{\hsize}{!}
 {\includegraphics[angle=-90]{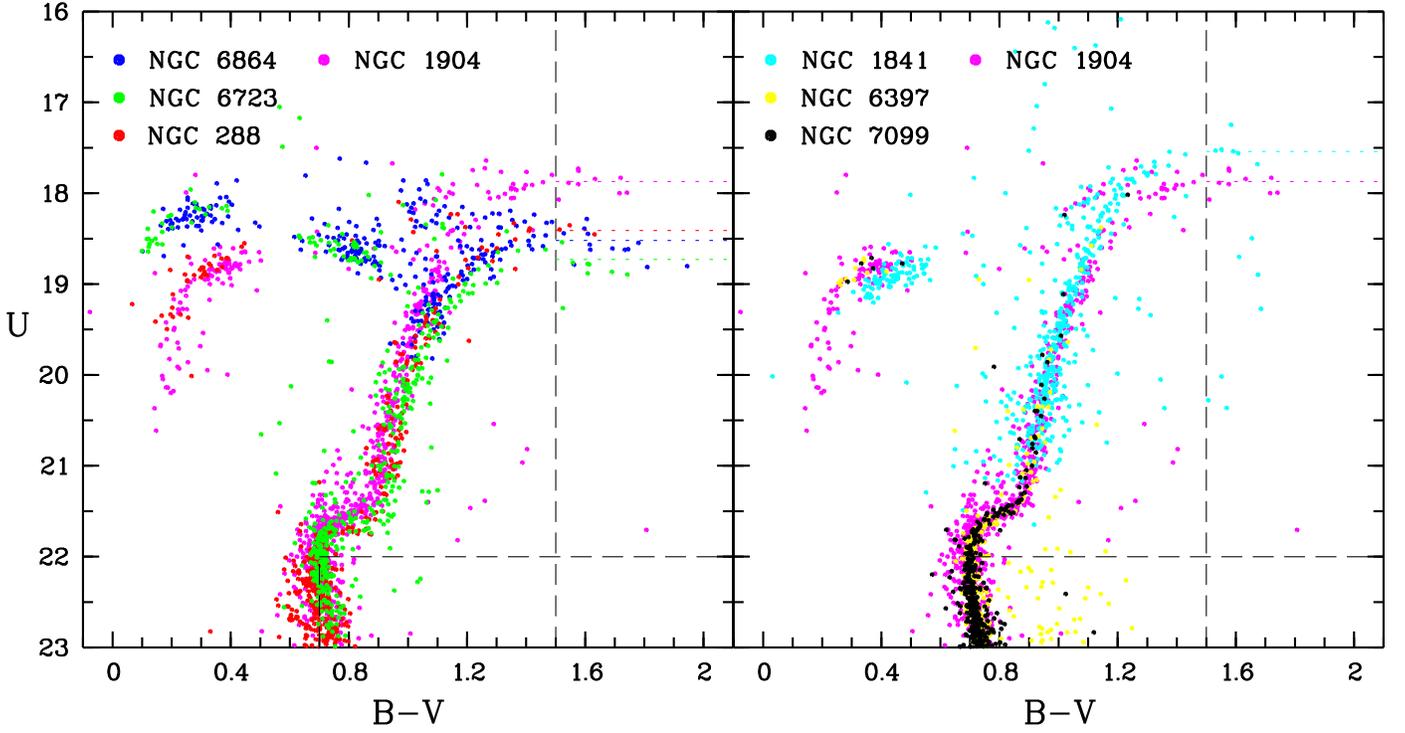}}
   \caption{Left panel: the superposition of the $U$($U\arcmin$)-$(B-V)$
diagrams of NGC 1904 (M79) and of more metal-rich GCs: NGC 288, NGC
6723 and NGC 6864 (M75); right panel: the superposition of the
$U\arcmin$-$(B-V)$ diagrams of NGC 1904 (M79) and of more metal-poor
GCs: NGC 1841, NGC 6397 and NGC 7099 (M30) the
$U$($U\arcmin$)-levels of inflection (brightest) point of the
cluster RGBs are shown by dashed lines of the same color as the
corresponding symbols of the cluster CMDs. Both the magnitudes and
colors of the original CMDs are reduced to those of the
corresponding CMD of M75.}
         \label{superpos3}
   \end{figure*}

To study the dependence of the $U$($U\arcmin$)-level of the RGB on
metallicity and to trace the BHB and define its position more
reliably in the $U\arcmin$-$(B-V)$ diagram (because of some
uncertainty in the position of the BHB of NGC 288 near the blue edge
of the instability strip) we used our photometry for a
sample of more metal-poor GCs. The magnitudes and colors of their
CMDs have been also reduced to those of the corresponding CMD of
M75. The data on the offsets applied to the original CMDs of all the
GCs, as well as the references to the sources of the original
photometric data are listed in Table~\ref{offset}. The superimposed
CMDs are presented in Fig.~\ref{superpos3}. Here, the location of BHB
in $U\arcmin$ is reliably and unambiguously shown by additional BHB
stars of a number of GCs with photometry reaching the TO points in
$U\arcmin$, namely of NGC 1904, NGC 6793, NGC 7099.

Finally, a result obtained several decades ago should be. In order to overcome
a number of disadvantages of the standard $U$ bandpass that are
related to the presence in it of the Balmer jump, Strai\v{z}ys
(\cite{straizys}) revised this bandpass and used a UV-filter with a
transmission curve blue-shifted of the jump. The resulting revised
UV bandpass, designated as "W", did not include the Bulmer jump. In
this sense, it is belongs to the
bandpasses conditionally denoted as $U\arcmin$. The
author showed that the relation between the color indexes
$(U-B)_{0}$ and $(W-B)_{0}$ was essentially nonlinear in a
certain range of the color $(B-V)_{0}$, with the maximum difference
between the former two indexes reaching around $(B-V)_{0}$ = 0, i.
e., just in the region of the BHB. These and many other data on
various photometric systems were summarized by Strai\v{z}ys
(\cite{straizys1}).
%%%%%%%%%%%%%%%%%%%%%%%%%%%%%%%%%%%%%%%%%%%%%%%%%%%%%%%%5
%
%%%%%%%%%%%%%%%%%%%%%%%%%%%%%%%%%%%%%%%%%%%%%%%%%%%%%%%%%%%%%%%%
%

\begin{table}
\begin{minipage}[t]{\columnwidth}
\caption{Data on the parameter $\Delta U_{TO}^{RGB}$ and
metallicities [Fe/H]$_{ZW}$ of GCs.} \label{uvparam} \centering
\renewcommand{\footnoterule}{}
\begin{tabular}{lcccc}
\hline \hline Cluster  &$\Delta U_{TO}^{RGB}$\footnote{The values
indicated in parentheses mean that cluster photometry does not reach
the main-sequence turnoff point} & $\sigma_{\Delta U}$ &
[Fe/H]$_{ZW}$ &
$\sigma_{[Fe/H]}$ \\
\hline
 NGC 288  & 3.59  & 0.10& -1.40 & 0.12\\
 NGC 1841  &(4.46) & 0.20  & -2.11& 0.15 \\
 NGC 1904  & 4.13  & 0.10   & -1.69  & 0.09 \\
 NGC 6723  & 3.27  & 0.10  & -1.09 & 0.14 \\
 NGC 6864  &(3.48)  & 0.20  & -1.32 & 0.12  \\
 \hline
\end{tabular}

\end{minipage}

\end{table}

\subsection{A sensitive indicator of metallicity in the CMDs with the UV magnitude axis}

Along with the bimodal appearance of the BHB in the CMD with the UV
magnitude axis, Fig.~\ref{superpos3} shows the position of the
$U$($U\arcmin$)-level of the upper RGB in GCs of different
metallicity. This level is defined unambiguously. It
corresponds to the inflection point of the RGB, the brightest
point of the branch in UV. Only AGB stars can produce some confusion
in the case of small number of brighter RGB stars. We defined the
parameter $\Delta U_{TO}^{RGB}$ as the difference in $U(U\arcmin)$
magnitude between the TO point and the inflection point of the RGB.
The color difference $\Delta (B-V)$ between these points is around
$\Delta (B-V) \approx$ 0.8 and almost independent of metallicity.
The mean location in the color of the inflection points of the RGBs is
shown by the vertical long-dashed line. Color dashed lines show the RGB
levels defined for the corresponding GC. The horizontal long-dashed
line is drawn at the level of the TO point of the superimposed CMDs.

The choice of the parameter $\Delta U_{TO}^{RGB}$ is clear. We
only note that in contrast to the $UV$-level of the HB (e.g.,
relative to that of the RGB), the $UV$-level of the TO point does
not apparently depend on the effect under consideration. This is
mainly due to the strong dependence of the Balmer jump on the
temperature. It decreases rapidly with decreasing temperature and
disappears (nearly) completely in stars of metal-poor GCs with
temperature close to that of the main-sequence TO point stars and
in cooler RGB stars. For this reason, $\Delta U_{TO}^{RGB}$
should not show a dependence, to a certain extent, on the
position of $UV$-bandpass relative the Balmer jump.

In our data base, among the most metal-poor GCs with available
photometry in UV, the Large Magellanic Cloud GC NGC 1841 is the only
GC with a large number of brighter RGB stars measured in this
bandpass. At the same time, the limiting magnitude of this
photometry does not reach the TO point of the cluster. For this
reason, as in the case of M75, we offset the CMD of NGC 1841 along
the $U\arcmin$-axis until coincidence of the mean $U\arcmin$-level
of its BHB and that of the BHBs of metal-poor GCs for which deep
photometry reaching the TO point in $U\arcmin$ is available (right
panel of Fig.~\ref{superpos3}). NGC 1841 is assumed to be as old as
the Galactic old metal-poor GCs. The offset along the
color axis was determined by the best coincidence of the lower RGB
of NGC 1841 and other most metal-poor GCs.

In Fig.~\ref{dependence}, we demonstrate, for the first time, the
dependence of the parameter $\Delta U_{TO}^{RGB}$ on cluster
metallicity, [Fe/H]$_{ZW}$, obtained for the set of metal-poor GCs.
Cluster metallicities (and the corresponding errors) are taken from
Zinn \& West (\cite {zinnwest}) for all the clusters, except NGC
1841. Its metallicity value is from Ferraro et al.
(\cite{ferraroetal}). The long-dashed line is a polynomial fit to the
data, assuming a linear relation between them. From this fit, we
derive the following simple relation between $\Delta U_{TO}^{RGB}$
and [Fe/H]$_{ZW}$ in the metal-poor range:

  \begin{equation}
    [Fe/H]_{\mathrm{ZW}} = 1.5 - 0.8\Delta U_\mathrm{{TO}}^{RGB} .
  \end{equation}

Fig.~\ref{dependence} shows a very tight correlation between these
values and the surprisingly small dispersion of the data. Note that it
becomes somewhat larger for the case of metallicity values derived
(for some clusters) from high-resolution spectroscopy or taken from
the Harris catalog (Harris {\cite{harris1}). The differences in
metallicity between M75, NGC 288, NGC 6723, estimated from the above
relation, using values of the parameter $\Delta U_{TO}^{RGB}$,
formally nearly coincide with those determined by relying on the
data from Zinn \& West (\cite {zinnwest}).

\begin{figure}
  \centering
 \includegraphics[angle=-90,width=8cm]{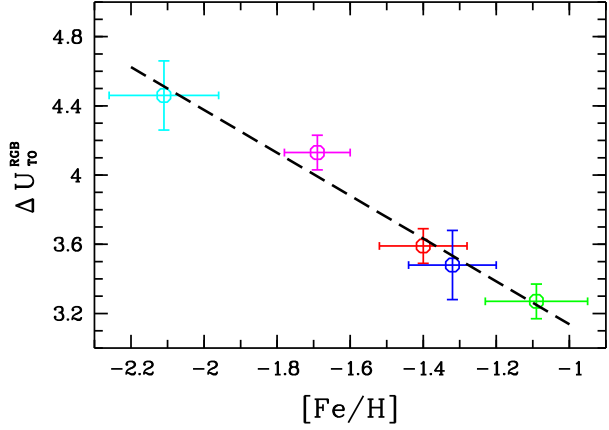}
\caption{The dependence of the parameter $\Delta U_{TO}^{RGB}$ on
cluster metallicity, [Fe/H]$_{ZW}$; long-dashed line is a polynomial
fit to the data, accepting a linear relation between them. Colors of
symbols and of error bars are the same GCs as in the previous
figure.}
         \label{dependence}
   \end{figure}

The parameter $\Delta U_{TO}^{RGB}$ is approximately twice as 
sensitive to metallicity variation than the position of the bump on
the RGB relative to the HB level in $V$ magnitude.

Table~\ref{uvparam} lists the GCs of our set, their metallicities,
estimated values of the parameter $\Delta U_{TO}^{RGB}$, as well as
errors of these characteristics.

Strictly speaking, the parameter $\Delta U_{TO}^{RGB}$ for a given
GC depends on the reddening of the cluster. Indeed, $\Delta
U_{TO}^{RGB}$ relies on $U$ magnitudes of stars belonging to
different luminosity classes and spectral types. In turn, extinction
of stars in any bandpass is a function of their temperature
(spectral type), luminosity class, and abundance (e.g., Grebel \&
Roberts \cite{grebrob}, and references therein). Therefore, the $\Delta
U_{TO}^{RGB}$ deduced is affected by the different dependence of
extinction A$_{U}$ of the RGB and main sequence TO stars on their
reddening $E(B-V)$. For this reason, we estimate the effect of
reddening on the values of $\Delta U_{TO}^{RGB}$ defined for the GCs
in our study. With this aim we used the calculations obtained in Grebel \& Roberts %%@
(\cite{grebrob}) and presented,
in particular, in their Tables 1 and 4, as well as in Figs. 4 and 5.
These calculations "produce ratios of the passband and colour
extinctions to A$_{V,0}$, the $V$ band extinction of a MS star with
a temperature of 17000K and solar metallicity". A$_{V,0}$ is defined
to be 1.0. According to the Harris catalog (Harris {\cite{harris1}),
NGC 1904 and NGC 6864 have the smallest [$E(B-V)=0.01$] and largest
[$E(B-V)=0.16$] reddening, respectively, among the GCs of our
sample. This means that the differences of the reddening and $V$
band extinction between the two GCs are $\Delta E(B-V)=0.15$ and
$\Delta$A$_{V}\approx$ 0.5, respectively. The difference in
metallicity between NGC 6864 and NGC 1904 is around $\Delta$[Fe/H]
$\approx$ 0.5 dex. The effective temperatures and surface gravities
for the stars at the main sequence TO and RGB inflection points are
accepted to be T$_{eff}\approx$ 6750K, $\log$ $g$ $\sim$ 5.0 and
T$_{eff}\approx$ 4250K, $\log$ $g$ $\sim$ 2.5. For
stars with such characteristics in the two clusters, we estimate
$\delta$[($\Delta U_{TO}^{RGB}$)$_{N1904}$ $-$ ($\Delta
U_{TO}^{RGB}$)$_{N6864}$] = ($\Delta$A$_{U,TO}$ $-$
$\Delta$A$_{U,RGB}$) $\sim$ 0.01 or so, where $\Delta$A$_{U,TO}$ is
the difference in the $U$ band extinction of main sequence TO stars
in NGC 6864 and in NGC 1904, and $\Delta$A$_{U,RGB}$ is the same for
stars at the RGB inflection points of the same GCs. The quantity
$\delta$ shows how many of the differences between $\Delta
U_{TO}^{RGB}$ of the two clusters would change if reddening of NGC
6864 was reduced to that of NGC 1904. It is at least an order of
magnitude lower than the typical error of the parameter under
consideration. Therefore, it cannot affect the dependence deduced
here between $\Delta U_{TO}^{RGB}$ and [Fe/H]. For the $U\arcmin$
bandpass, the above-estimated effect is of the same order of
magnitude, i.e. it is small compared to the the typical error of
$\Delta U_{TO}^{RGB}$.

The impact of differences in the response curves of differently
realized $U$ bandpass on the dependence of the corresponding extinction
A$_{U}$ on reddening $E(B-V)$ is another factor that can
affect relative positions of stars in $U$ magnitude, in
particular $\Delta U_{TO}^{RGB}$ or $\Delta U_{TO}^{BHB}$, the $U$
magnitude difference between the main sequence TO and BHB. The larger variations of the %%@
difference between extinction
A$_{U\arcmin}$ and A$_{U}$ as a function of $E(B-V)$ among stars
belonging to distinct spectral types and luminosity classes, the
larger the variations between $\Delta U$ and $\Delta U\arcmin$, defined
for these stars, as a function of reddening. In other words, if the
difference between A$_{U\arcmin}$[$E(B-V)$] and A$_{U}$[$E(B-V)$]
was the same for stars of any spectral type and luminosity class,
the impact of the factor under consideration would be zero. As an
example, we compare the difference between
A$_{U\arcmin}$[$E(B-V)$] and A$_{U}$[$E(B-V)$] for the ultraviolet
bandpass realized with the F336W filter at HST and the standard $U$
bandpass, respectively. The transmission curve of F336W (e.g.,
Holtzman et al. \cite{holtzmanetal}) is notably blue-shifted of the
Bulmer jump. So, the corresponding ultraviolet bandpass is an
example of the conditionally defined $U\arcmin$ bandpass. Holtzman
et al. (\cite{holtzmanetal}) have presented in their Figure 12 "a
plot of extinction for several WFPC2 filters and for $UBVRI$ as
function of $E(B-V)$ up to $E(B-V)$=1.0", using synthetic
computations with O6 and K5 input spectra. In turn, Tables 12a
and 12b from the paper contain the same data, but for WFPC2 filters
alone. The difference between extinction
A$_{F336W}$ and A$_{U}$ is insignificant at low reddening for the
early spectral type O6. It is of the order of a few
hundredth of a mag at $E(B-V)=0.15$. For late spectral types K5,
however, this difference is larger with opposite
sign. Apart from this very approximate evaluation we obtained more
rigorous estimates. We calculated A$_{U}$ as a function of $E(B-V)$,
using analytical expressions deduced by A\v{z}usienis \&
Strai\v{z}ys (\cite{azustraiz}), and data from their Tables 1 and 2.
We found that A$_{F336W}$(O6) $-$ A$_{U}$(O7) $\approx$ 0.03 at
$E(B-V)=0.15$ for the early stars. For the late spectral type K5 and
at the same reddening, A$_{F336W}$(K5) $-$ A$_{U}$(K5) $\approx
-0.18$. These results are in reasonable agreement with what can be
estimated from the graphically presented dependence of the
extinction A$_{U}$ and A$_{F336W}$ on $E(B-V)$ by Holtzman et al.
(\cite{holtzmanetal}). This change of the dependence of the
extinction A$_{F336W}$ on $E(B-V)$ for the K5 spectral type is
mainly due to the fairly significant red leak of F336W. The red leak
slightly affects the dependence of the $U$ ($U\arcmin$) band
extinction on $E(B-V)$ for early stars, but its impact can be
significant in the case of late stars, especially with large amounts of
absorbing matter (e.g., A\v{z}usienis \& Strai\v{z}ys
\cite{azustraiz}). Therefore, when reddening increases, the
parameter $\Delta U_{TO}^{RGB}$ determined from observations with an
ultraviolet filter having (significant) red leak is expected to have
increasing systematic error (comparable to the random error or
exceeding it), as compared to the same parameter obtained using $U$
or $U\arcmin$ filters without red leak. The same effect (but somewhat less
strong and with opposite sign) should affect the magnitude difference 
$\Delta U_{TO}^{BHB}$, as well.

\section{Conclusions}

We obtained new multi-color photometry of more than 2600 stars in
a wide field of the southern GC M75, above its turnoff point. We used
this photometry to analyze the brighter sequences of CMD.

We were able to isolate, for the first time, the AGB clump and to
determine the parameter $\Delta V_{HB}^{clump}$ = 1.07 $\pm$ 0.06,
the difference in $V$ magnitude between the ZAHB level and that of
the clump. Also, by relying on the obtained LF of the upper RGB, we
estimated similar parameter for the RGB bump, $\Delta V_{HB}^{bump}$
= 0.07 $\pm$ 0.08, as well as the behavior and features of
the LF.

We combine the present $U$ photometry of M75 with our analogous
photometric data on metal-poor GCs, [Fe/H]$_{ZW} < -1.1$ dex,
obtained from observations with different equipment, but calibrated
by standard stars located in the observed cluster fields. We
investigate the position of the HB in deep CMDs reaching the turnoff point,
with UV magnitude axis. We demonstrate that the difference in
$U$ magnitude, $\Delta U_{TO}^{BHB}$, between the TO point and the
$U$-level of the blue HB near the blue edge of the instability strip
is bimodal. The BHB of GCs observed with the EMMI camera of the
NTT/ESO telescope is about 0.4 - 0.5 mag smaller as compared to GCs
observed with the 100$\arcsec$ telescope and 1.3 m Warsaw telescope
of the Las Campanas Observatory. However, another parameter, $\Delta
U_{TO}^{RGB}$, the difference in $U$ magnitude between the TO point
and inflection point of the RGB, does not obviously depend on the
characteristics of $U$ filters used, at least within insignificant
reddening and provided that the filters do not have significant red
leak. It shows very tight, nearly perfect correlation with cluster
metallicities, [Fe/H]$_{ZW}$, taken from Zinn \& West (\cite
{zinnwest}). We deduce, for the first time, an analytical relation
between $\Delta U_{TO}^{RGB}$ and [Fe/H]$_{ZW}$.

From the point of view of practical
application of the demonstrated sensitive indicator of metallicity,
especially in the case of photometry of resolved stars in GCs
populating near galaxies, one must apply a
slightly modified approach. Specifically, the parameter $\Delta
U_{BHB}^{RGB}$ (the difference in $U$ magnitude between the BHB
level and inflection point of the RGB) is more easily measurable
than $\Delta U_{TO}^{RGB}$. Moreover, the former is presumably less
dependent on cluster age than the latter. Formally, it is quite easy
to transform the obtained relation between $\Delta U_{TO}^{RGB}$ and
[Fe/H]$_{ZW}$ into the relation between $\Delta U_{BHB}^{RGB}$ and
[Fe/H]$_{ZW}$ by accepting the same slope for both relations. For
this purpose, it is sufficient to take into account the difference
in $U$($U\arcmin$) magnitude between the TO point and $U$-level of
the BHB: $\Delta U_{BHB}^{RGB}$ = $\Delta U_{TO}^{RGB} -$ 3.8  and
$\Delta U\arcmin_{BHB}^{RGB}$ = $\Delta U_{TO}^{RGB} -$ 3.3 for the
$U$ and $U\arcmin$ bandpasses, respectively. Here, the $U$-level of
the BHB is accepted in the same sense as proposed by Ferraro et al.
(\cite{ferraroetal}) and used by us above to define the $V$-level of the
HB. However, these estimates are very
preliminary. To achieve more reliable conclusions, more
observational data on other GCs in the metal-poor range are
required. In analogy with $V$-level of the BHB, one can
expect that it is difficult to reliably define the $U$-level of the BHB
for some GCs, particularly for those clusters, like M13, that
exhibit both pure BHB morphology and an extended BHB tail. Note also
that due to its larger color base the parameter $\Delta U_{BHB}^{RGB}$ 
is probably more affected by red leak than the parameter 
$\Delta U_{TO}^{RGB}$.

The obtained results concern metal-poor GCs. Deep UV photometry in
metal-rich GCs, [Fe/H]$_{ZW} > -1.0$, is needed to investigate the
same problems in this range of metallicity. Moreover, strictly
speaking, in our sample of GCs observed at NTT (i.e. in $U\arcmin$
bandpass) there is no GC with sufficient number of stars belonging
to the RHB. Thus it is difficult to reach definite
conclusions about the apparent position (relative to the TO point and
BHB) of the RHB in CMDs with $U$ and $U\arcmin$ magnitude axes.
Also, it is unclear whether (and how exactly) the RHB
position in these diagrams depends on metallicity. Our very preliminary
conclusion, based on indirect and insufficient evidence is
that the position of the RHB, like that of the RGB, does not show a
(notable) dependence, as does BHB, on the presence or absence of the
Bulmer jump in UV ($U$ or $U\arcmin$) bandpass.

\begin{acknowledgements}
We thank the anonymous referee for useful comments that have
improved the manuscript. We are grateful to Marcio Catelan and
Andrzej Udalski for kindly providing data on photometry in M75 and on the
response curve of the $U$ bandpass realized at the 1.3 m Warsaw
telescope. VK acknowledges support from the
Universidad Cat\'olica del Norte through research grant DGIP
10301180.

\end{acknowledgements}

\end{document}